\documentclass[a4paper,11pt]{article}
\usepackage[margin=2.5cm]{geometry}
\usepackage{lineno}
\usepackage{amsfonts,amsmath,amssymb}
\usepackage[titletoc,toc,title]{appendix}
\usepackage{authblk}
\usepackage{braket}
\usepackage{cancel}
\usepackage[font={footnotesize,it}]{caption}
\usepackage{cite}
\usepackage{comment}
\usepackage{enumitem}
\usepackage{epsfig}
\usepackage{etoolbox}
\usepackage{float}
\usepackage{graphicx}
\usepackage{ragged2e}
\usepackage{setspace}
\usepackage{soul}
\usepackage{subcaption}
\usepackage{hyperref}%after other packages
\usepackage{cleveref}%after hyperref
\newcommand*\linenomathpatch[1]{%
  \cspreto{#1}{\linenomath}%
  \cspreto{#1*}{\linenomath}%
  \csappto{end#1}{\endlinenomath}%
  \csappto{end#1*}{\endlinenomath}%
}
\linenomathpatch{equation}
\linenomathpatch{align}
\hypersetup{colorlinks=true,citecolor=red,linkcolor=blue,urlcolor=blue,linktocpage=true}
{\makeatletter\g@addto@macro\bfseries{\boldmath}\makeatother}
\numberwithin{equation}{section}
\crefname{subsection}{subsection}{subsections}
\def\cft#1{CFT$_{#1}$}
\def\ads#1{AdS$_{#1}$}
\def\renyi{R{\'e}nyi}
\def\poincare{Poincar{\'e}}
\begin{document}
%%%%%
\title{On minimal entanglement wedge cross section for holographic entanglement negativity}
%%%%%
\author[1,2]{Jaydeep Kumar Basak\thanks{E-mail: \texttt{jaydeep@mail.nsysu.edu.tw}}}
\author[3]{Vinay Malvimat\thanks{E-mail: \texttt{vinay.malvimat@apctp.org}}}
\author[4,5]{Himanshu Parihar\thanks{E-mail: \texttt{himansp@phys.ncts.ntu.edu.tw}}}
\author[6]{Boudhayan Paul\thanks{E-mail: \texttt{paul@iitk.ac.in}}}
\author[6]{Gautam Sengupta\thanks{E-mail: \texttt{sengupta@iitk.ac.in}}}
%%%%%
\affil[1]{Department of Physics, National Sun Yat-Sen University, Kaohsiung 80424, Taiwan
\medskip}
%%%%%
\affil[2]{Center for Theoretical and Computational Physics, Kaohsiung 80424, Taiwan
\medskip}
%%%%%
\affil[3]{Asia Pacific Center for Theoretical Physics, Pohang 37673, Korea
\medskip}
%%%%%
\affil[4]{Center of Theory and Computation, National Tsing-Hua University, Hsinchu 30013, Taiwan
\medskip}
%%%%%
\affil[5]{Physics Division, National Center for Theoretical Sciences, Taipei 10617, Taiwan
\medskip}
%%%%%
\affil[6]{Department of Physics, Indian Institute of Technology, Kanpur 208016, India}
%%%%%
\date{}
\maketitle
%%%%%
\begin{abstract}
We demonstrate the equivalence of two different conjectures in the literature for the holographic entanglement negativity in AdS$_3$/CFT$_2$, modulo certain constants. These proposals involve certain algebraic sums of bulk geodesics homologous to specific combinations of subsystems, and the entanglement wedge cross section (EWCS) backreacted by a cosmic brane for the conical defect geometry in the bulk gravitational path integral. It is observed that the former conjectures reproduce the field theory replica technique results in the large central charge limit whereas the latter involves constants related to the Markov gap. In this context we establish an alternative construction for the EWCS of a single interval in a CFT$_2$ at a finite temperature to resolve an issue for the latter proposal involving the thermal entropy elimination for the holographic entanglement negativity. Our construction for the EWCS correctly reproduces the corresponding field theory results modulo the Markov gap constant in the large central charge limit.
\end{abstract}
%%%%%
\clearpage
\tableofcontents
\clearpage
%%%%%

\section{Introduction}\label{sec_intro}

Quantum entanglement has evolved as one of the dominant themes in diverse disciplines from many body condensed matter systems to black holes and quantum gravity. In this context the entanglement entropy (EE), defined as the von Neumann entropy of the reduced density matrix has been central to the characterization of pure state entanglement. For mixed states however, the EE fails to correctly capture the entanglement as it involves irrelevant classical and quantum correlations (\textit{e.g.}, for finite temperature configurations, it includes the thermal correlations). Hence the characterization of mixed state entanglement has been a significant issue in quantum information theory leading to the proposals of various entanglement and correlation measures in the recent past. In this connection Vidal and Werner \cite{Vidal} proposed a computable mixed state entanglement measure termed entanglement (logarithmic) negativity (EN) which characterized an upper bound on the distillable entanglement.\footnote{Several other measures to characterize mixed state entanglement have also been proposed in quantum information theory. However most of these involve optimization over LOCC protocols and are not directly computable.} It was defined as the logarithm of the trace norm for the partially transposed reduced density matrix. Subsequently Plenio \cite{Plenio:2005cwa} established that despite being non convex, the entanglement negativity was an entanglement monotone under local operations and classical communication (LOCC) which justified its utility for the characterization of mixed state entanglement.

In a series of communications \cite{Calabrese:2004eu,Calabrese:2009qy,Calabrese:2009ez,Calabrese:2010he} the authors formulated a replica technique to compute the entanglement entropy in two dimensional conformal field theories (\cft{2}s). The procedure was later extended to configurations with multiple disjoint intervals in \cite{Hartman:2013mia,Headrick:2010zt}, where it was shown that the entanglement entropy receives non universal contributions depending on the full operator content of the theory which were sub leading in the large central charge limit. A variant of the above replica technique was developed in \cite{Calabrese:2012ew,Calabrese:2012nk,Calabrese:2014yza} to compute the entanglement negativity of bipartite states in \cft{2}s. It was also shown in \cite{Kulaxizi:2014nma} that for the mixed state of two disjoint intervals the entanglement negativity was non universal, however it was possible to isolate a universal contribution in the large central charge limit. Interestingly the entanglement negativity for this configuration was numerically shown to exhibit a phase transition depending upon the separation of the two intervals \cite{Kulaxizi:2014nma,Dong:2018esp}.

In a major development Ryu and Takayanagi (RT) \cite{Ryu:2006bv,Ryu:2006ef} proposed a holographic conjecture for the EE of a subsystem in a dual \cft{d} involving the area of a homologous bulk codimension-two static minimal surface (RT surface), in the context of the \ads{d+1}/\cft{d} correspondence. This significant proposal led to the emergence of the field of holographic quantum entanglement (for a detailed review see \cite{Ryu:2006ef,Nishioka:2009un,Rangamani:2016dms,Nishioka:2018khk}). The RT conjecture was proved initially for the \ads{3}/\cft{2} scenario, with later generalization to the \ads{d+1}/\cft{d} framework in \cite{Fursaev:2006ih,Casini:2011kv,Faulkner:2013yia,Lewkowycz:2013nqa}. Hubeny, Rangamani and Takayanagi (HRT) extended the RT conjecture to covariant scenarios in \cite{Hubeny:2007xt}, a proof of which was established in \cite{Dong:2016hjy}.

Naturally the above developments motivated the investigation of a corresponding holographic characterization for the entanglement negativity. One of the first steps in this direction was proposed in \cite{Rangamani:2014ywa} for the pure vacuum state of a \cft{d} dual to a bulk pure \ads{d+1} geometry although a general prescription for arbitrary bipartite states remained elusive. This significant issue was addressed in \cite{Chaturvedi:2016rcn,Chaturvedi:2016opa} where a holographic entanglement negativity conjecture and its covariant extension were advanced for bipartite mixed state configurations in the \ads{3}/\cft{2} scenario, with the generalization to higher dimensions reported in \cite{Chaturvedi:2016rft}. These proposals involved certain algebraic sums of bulk geodesics homologous to appropriate combinations of subsystems. A large central charge analysis of the entanglement negativity through the monodromy technique for holographic \cft{2}s was established in \cite{Malvimat:2017yaj} which provided a strong substantiation for the proposals described above. Subsequently in a series of works the above holographic conjectures and their covariant extensions were utilized to obtain the entanglement negativity of various bipartite states in \cft{2}s and their higher dimensional generalizations \cite{Jain:2017aqk,Jain:2017uhe,Malvimat:2018txq,Malvimat:2018ood,Basak:2020bot}.

On a different note, motivated by the quantum error correcting codes, an alternate approach involving the backreacted entanglement wedge cross section (EWCS) to compute the holographic entanglement negativity for configurations with spherical entangling surfaces was advanced in \cite{Kudler-Flam:2018qjo}. Furthermore a proof for this proposal, based on the reflected entropy \cite{Dutta:2019gen} was established in another communication \cite{Kusuki:2019zsp}. The entanglement wedge was earlier shown to be the bulk subregion dual to the reduced density matrix of the dual \cft{}s in \cite{Czech:2012bh,Wall:2012uf,Headrick:2014cta,Jafferis:2014lza,Jafferis:2015del}. Also the EWCS has been proposed to be the bulk dual of the entanglement of purification (EoP) \cite{Takayanagi:2017knl,Nguyen:2017yqw} (for recent progress see \cite{Terhal:2002riz,Bhattacharyya:2018sbw,Bao:2017nhh,Hirai:2018jwy,Espindola:2018ozt,Umemoto:2018jpc,Bao:2018gck,Umemoto:2019jlz,Guo:2019pfl,Bao:2019wcf,Harper:2019lff}). The connection of the EWCS to the odd entanglement entropy (OEE) \cite{Tamaoka:2018ned} and the reflected entropy \cite{Dutta:2019gen,Jeong:2019xdr,Bao:2019zqc,Chu:2019etd} has also been explored.

As mentioned above, in \cite{Kudler-Flam:2018qjo,Kusuki:2019zsp} the authors proposed that for configurations involving spherical entangling surfaces, the holographic entanglement negativity may be expressed in terms of the EWCS backreacted by a cosmic brane for the conical defect of the replicated bulk geometry in a gravitational path integral. Utilizing this conjecture the authors computed the holographic entanglement negativity for bipartite states in \cft{2}s dual to bulk pure \ads{3} geometries and planar BTZ black hole, through the construction described in \cite{Takayanagi:2017knl}. Their results for the holographic entanglement negativity following from the above construction reproduced the corresponding field theory replica technique results in the large central charge limit, up to certain constants involving the Markov gap between the reflected entropy and the mutual information \cite{Hayden:2021gno}, except for the configuration of a single interval in a finite temperature \cft{2} described in \cite{Calabrese:2014yza}. Specifically their result for this mixed state configuration missed the subtracted thermal entropy term in the expression for the entanglement negativity \cite{Calabrese:2014yza}. Given that their construction exactly reproduces the replica technique results for all the other mixed state configurations, the mismatch described above requires a resolution.

In this article we demonstrate the equivalence of the two holographic proposals up to constants involving the Markov gap. In this connection we also address the intriguing issue of the missing thermal entropy term for the single interval configuration at a finite temperature for the second proposal based on the EWCS through an alternative construction. Our construction is inspired by that of Calabrese \textit{et al.\@} \cite{Calabrese:2014yza} and involves two symmetric auxiliary intervals on either side of the single interval under consideration. In this construction we have utilized certain properties of the EWCS along with a specific relation valid for the dual bulk BTZ black hole geometry. Finally we implement the bipartite limit where the auxiliary intervals are allowed to be infinite and constitute the rest of the system, to arrive at the correct minimal EWCS for the configuration in question. The holographic entanglement negativity computed using the conjecture advanced in \cite{Kudler-Flam:2018qjo,Kusuki:2019zsp} through our alternative construction for the EWCS correctly reproduces the corresponding replica result in \cite{Calabrese:2014yza} mentioned earlier. In particular we are able to obtain the missing thermal entropy term in the expression for the holographic entanglement negativity described in \cite{Calabrese:2014yza}. We further observe that the monodromy analysis employed by the authors in \cite{Kusuki:2019zsp} and that in \cite{Malvimat:2017yaj} lead to identical functional forms for the relevant four point correlation function for the twist fields, in the large central charge limit for the mixed state configuration of the single interval at a finite temperature.

This article is organized as follows. In \cref{sec_hen_review} we briefly review the definition and holographic constructions involving the bulk geodesics for the entanglement negativity in the \ads{3}/\cft{2} scenario. In \cref{sec_ew_review} we review the entanglement wedge construction in \cite{Takayanagi:2017knl}. In \cref{sn_HEN_EWCS} we describe the computation of the holographic entanglement negativity based on the EWCS in \cite{Kudler-Flam:2018qjo,Kusuki:2019zsp} and demonstrate the equivalence of their results with those obtained from the former proposal. Following this in \cref{sn_issues} we describe an issue with the holographic entanglement negativity for a single interval at a finite temperature obtained from the EWCS, involving a subtracted thermal entropy term. We further propose an alternative construction of the EWCS for this configuration which resolves this issue and restores the missing thermal entropy term. Finally, we summarize our results in \cref{sec_summary} and present our conclusions. In \cref{app_hendj} we have included a short review of the derivation for the entanglement negativity of two disjoint intervals in a \cft{2}. Additionally in \cref{app_proof,app_gap} we briefly describe a sketch of a plausible proof of the holographic entanglement negativity proposal based on bulk geodesics from a gravitational path integral perspective, and the issue of the holographic Markov gap.

\section{Review of earlier literature}\label{sec_hen_review}

In this section we review the holographic proposals for the entanglement negativity involving certain algebraic sums of the holographic \renyi{} entropies of order half described by the lengths of backreacting cosmic branes homologous to the subsystems as described in \cite{KumarBasak:2020ams}.

\subsection{\renyi{} entanglement entropy}\label{HREE}

Here we briefly recapitulate the holographic construction for the \renyi{} entropy which was proposed in \cite{Dong:2016fnf}. The author in \cite{Dong:2016fnf} utilized the gravitational path integral technique developed in \cite{Lewkowycz:2013nqa} to demonstrate that the holographic \renyi{} entropy of a subsystem in a \cft{} is related to the area of a codimension-two cosmic brane $\mathcal{C}_n$ with tension $T_n=\frac{n-1}{4nG_N}$ in the replicated bulk geometry, homologous to the subsystem under consideration as
\begin{align}
n^{2} \frac{\partial}{\partial n}\left(\frac{n-1}{n}S^{(n)}(A)\right)
&=\frac{\text{Area}\left(\mathcal{C}_n\right)}{4G_N},\label{AreaCB1}\\
n^{2} \frac{\partial}{\partial n}\left(\frac{n-1}{n} \mathcal{A}^{(n)}\right)
&=\text{Area}\left(\mathcal{C}_n\right),\label{AreaCB}
\end{align}
where $S^{(n)}(A)$ is the \renyi{} entanglement entropy of order $n$ for the subsystem $A$. Note that $\mathcal{A}^{(n)}$ is related to $S^{(n)}$ as follows
\begin{equation}\label{Xidong}
S^{(n)}=\frac{\mathcal{A}^{(n)}}{4G_N}.
\end{equation}
In the replica limit $n\to 1$, the tension vanishes and \cref{AreaCB1,AreaCB,Xidong} reduce to that of the usual RT proposal as
\begin{equation}\label{rt}
S_Y=\frac{\mathcal{A}_Y}{4G_N},
\end{equation}
where $\mathcal{A}_{Y}$ denotes the area of the minimal surface homologous to the subsystem $Y$ and $G_N$ is the $d+1$ dimensional gravitational constant. However, for $n\neq 1$ the backreaction from the brane is non-zero and it is difficult to determine the area of a cosmic brane homologous to a subsystem of an arbitrary geometry. Interestingly for subsystems with spherical entangling surfaces the backreaction was explicitly computed\footnote{The authors in \cite{Hung:2011nu} utilized a conformal map from a hyperbolic cylinder to the causal evolution of a subregion enclosed by a spherical entangling surface in flat space. This in turn implies that the entanglement entropy of a spherical region in a \cft{} on a flat Minkowski space is given by an integral of thermal entropy of a \cft{} on a hyperbolic cylinder. In the context of \ads{}/\cft{} correspondence this translates to computing the horizon entropy of a certain topological black hole by the well known Wald formula.} in \cite{Hung:2011nu} where the holographic \renyi{} entanglement entropy of a subsystem $A$ in a \cft{d} was expressed as
\begin{equation}\label{AreaRel}
S^{(n)}(A)=\mathcal{X}^{(n)}_d \, S(A).
\end{equation}
In \cref{AreaRel}, the proportionality constant has the following form
\begin{equation}\label{chid}
\mathcal{X}^{(n)}_d=\frac{n}{2(n-1)}\left(2-x_n^{d-2}\left(1+x_n^2\right)\right),
\end{equation}
where
\begin{equation}\label{xn}
x_n=\frac{1}{nd}\left(1+\sqrt{1-2dn^2+d^2n^2}\right).
\end{equation}
In the \ads{3}/\cft{2} scenario, the holographic \renyi{} entropy of order half is given as
\begin{align}\label{SAchi}
S^{(1/2)}(A)&=\mathcal{X}^{(1/2)}_d \, S(A)\nonumber\\
&=\frac{3}{2}S(A)\nonumber\\
&=\frac{3}{8G_N}\mathcal{L}_A,
\end{align}
where in the second line we have used \cref{chid,xn} to determine the constant $\mathcal{X}_2=3/2$ 
and $\mathcal{L}_A$ denotes the length of the geodesic homologous to the subsystem $A$.

\subsection{Entanglement negativity (EN) in a \cft{2}}\label{EN_CFT}

In this subsection we provide a brief review of the definition of the entanglement negativity and its computation in a \cft{2}. In this connection we consider a tripartite system $ABC$ in a pure state, comprising the subsystems $A$, $B$ and $C$. The reduced density matrix $\rho_{AB}$ for the mixed state bipartite configuration $AB\equiv A\cup B$ may then be obtained by tracing over the subsystem $C$. The relevant Hilbert space is given by $\mathcal{H}_{AB}=\mathcal{H}_A\otimes\mathcal{H}_B$, where $\mathcal{H}_{A,B}$ represents the Hilbert space for the subsystem $A,B$ respectively. The partial transpose of the reduced density matrix $\rho_{AB}$, denoted by $\rho_{AB}^{T_B}$, may then be defined as follows
\begin{equation}\label{pt_def}
\left\langle e_i^{(A)}e_j^{(B)}\middle\lvert\rho_{AB}^{T_B}\middle\rvert e_k^{(A)}e_l^{(B)}\right\rangle
=\left\langle e_i^{(A)}e_l^{(B)}\middle\lvert\rho_{A B}\middle\rvert e_k^{(A)}e_j^{(B)}\right\rangle,
\end{equation}
where $\lvert e_i^{(A,B)}\rangle$ describes the basis for $\mathcal{H}_{A,B}$ respectively. The entanglement negativity between the subsystems $A$ and $B$ is then defined in terms of the trace norm\footnote{The trace norm for an arbitrary hermitian matrix $M$ is given by $\lVert M\rVert= \text{Tr}\left(\sqrt{MM^\dagger}\right)$.} of the partially transposed reduced density matrix $\rho_{AB}^{T_B}$ as follows
\begin{equation}\label{EN_def}
\mathcal{E}(A:B)=\log\left\lVert\rho_{AB}^{T_B}\right\rVert.
\end{equation}

In this context we introduce below the definition of the \textit{\renyi{} entanglement negativity} (REN) of order $k$ following \cite{Calabrese:2012ew,Calabrese:2012nk} as
\begin{equation}\label{REN_def}
\mathcal{N}^{(k)}(A:B)=\text{Tr}\left[\left(\rho_{AB}^{T_B}\right)^k\right],
\end{equation}
where $k$ is a positive integer. The EN as defined in \cref{EN_def} may then be expressed in terms of the REN as
\begin{equation}\label{EN_REN_def}
\mathcal{E}(A:B)=\lim_{n_e\to 1}\log\left[\mathcal{N}^{(n_e)}(A:B)\right]
=\lim_{n_e\to 1}\log\left[\text{Tr}\left(\rho_{AB}^{T_B}\right)^{n_e}\right].
\end{equation}
Note that the right hand side of \cref{EN_REN_def} indicates the analytic continuation of the REN of even orders, denoted by $n_e\in 2\mathbb{Z}^+$, to $n_e=1$.

For a \cft{2}, the authors in \cite{Calabrese:2012ew,Calabrese:2012nk,Calabrese:2014yza} developed a replica technique involving $n_e\in 2\mathbb{Z^+}$ replicas of the original complex manifold $\mathcal{M}$ with branch cuts along $A$ and $B$. The trace in \cref{EN_REN_def} may then be evaluated in terms of certain twist field correlators involving the end points of $A$ and $B$. For example, in a zero temperature \cft{2}, when the subsystems (intervals in a \cft{2}) $A=[a_1,a_2]$ and $B=[b_1,b_2]$ are disjoint, the entanglement negativity is given by the following four point twist correlator
\begin{equation}
\label{EN_twist}
\mathcal{E}(A:B)=\lim_{n_e\to 1}\log
\left\langle\mathcal{T}_{n_e}(a_1)\overline{\mathcal{T}}_{n_e}(a_2)
\overline{\mathcal{T}}_{n_e}(b_1)\mathcal{T}_{n_e}(b_2)\right\rangle,
\end{equation}
where $\mathcal{T}_{n_e}$ and $\overline{\mathcal{T}}_{n_e}$ are twist and anti-twist operators respectively.

\subsection{Holographic entanglement negativity (HEN)}\label{intro-hen}

As discussed earlier, a replica technique proposed in \cite{Calabrese:2012ew,Calabrese:2012nk} was utilized to compute the entanglement negativity for various pure and mixed state configurations of a \cft{2}. Following this, several holographic proposals for the entanglement negativity of various configurations were advanced in \cite{Chaturvedi:2016rcn,Jain:2017aqk,Malvimat:2018txq} in terms of appropriate algebraic sums of the lengths of bulk geodesics in the \ads{3}/\cft{2} framework. For example, the holographic entanglement negativity of two disjoint intervals $A$ and $B$ in proximity is given by \cite{Malvimat:2018txq}
\begin{align}
\mathcal{E}&=\left.\frac{3}{16G_N}\middle[\mathcal{L}_{A\cup C}+\mathcal{L}_{B\cup C}-\mathcal{L}_{A\cup B\cup C}-\mathcal{L}_{C}\right]\label{DJArea}\\
&=\left.\frac{3}{4}\middle[S(A\cup C)+S(B\cup C)-S(A\cup B\cup C)-S(C)\right],
\end{align}
where $C$ denotes the interval sandwiched between $A$ and $B$. From the discussion in \cref{HREE}, it is clear that we may re-express the conjecture given in \cref{DJArea} as follows
\begin{equation}
\mathcal{E}=\left.\frac{\mathcal{X}_2}{8G_N}\middle[\mathcal{L}_{A\cup C}+\mathcal{L}_{B\cup C}-\mathcal{L}_{A\cup B\cup C}-\mathcal{L}_{C}\right].\nonumber
\end{equation}
In the \ads{3}/\cft{2} scenario, we now utilize the result given in \cref{SAchi} to rewrite the above expression in the following way
\begin{align}
\mathcal{E}&=\frac{1}{8G_N}\left[\mathcal{L}^{(1/2)}_{A\cup C}+\mathcal{L}^{(1/2)}_{B\cup C}
-\mathcal{L}^{(1/2)}_{A\cup B\cup C}-\mathcal{L}^{(1/2)}_{ C}\right]\\
&=\frac{1}{2}\left[S^{(1/2)}(A\cup C)+S^{(1/2)}(B\cup C)-S^{(1/2)}(A\cup B\cup C)-S^{(1/2)}(C)\right].\label{HENDJ}
\end{align}
Following the same procedure as above we may re-express the holographic conjecture for the entanglement negativity of two adjacent intervals in \cite{Jain:2017aqk} as
\begin{equation}
\mathcal{E}=\frac{1}{2}\left[S^{(1/2)}(A)+S^{(1/2)}(B)-S^{(1/2)}(A\cup B)\right]\label{HENADJ}.
\end{equation}
Similarly, the holographic conjecture for the entanglement negativity of a single interval described in \cite{Chaturvedi:2016rcn} may be expressed as follows
\begin{equation}
\mathcal{E}=\lim_{B_1\cup B_2\to A^c}\left.\frac{1}{2}\middle[2S^{(1/2)}(A)+S^{(1/2)}(B_1)+S^{(1/2)}(B_2)
-S^{(1/2)}(A\cup B_1)-S^{(1/2)}(A\cup B_2)\right].\label{HENSIN}
\end{equation}
A plausible derivation of these proposals as expressed in \cref{HENDJ,HENADJ,HENSIN}, based on the replica symmetry breaking saddles for a gravitational path integral for spherical entangling surfaces, is described briefly in \cref{app_proof}.\footnote{For details of this proof refer to \cite{KumarBasak:2020ams}.}

As mentioned earlier in the \nameref{sec_intro}, the above holographic proposals for the entanglement negativity were further generalized to covariant scenarios in \cite{Chaturvedi:2016opa,Jain:2017uhe,Malvimat:2018ood}. Additionally these holographic proposals were also extended to the general \ads{d+1}/\cft{d} framework in terms of similar algebraic sums of the areas of bulk RT surfaces for certain combinations of the subsystems \cite{Chaturvedi:2016rft,Basak:2020bot}.

\section{Entanglement wedge cross section (EWCS)}\label{sec_ew_review}

\begin{figure}
\centering
\includegraphics[scale=.45]{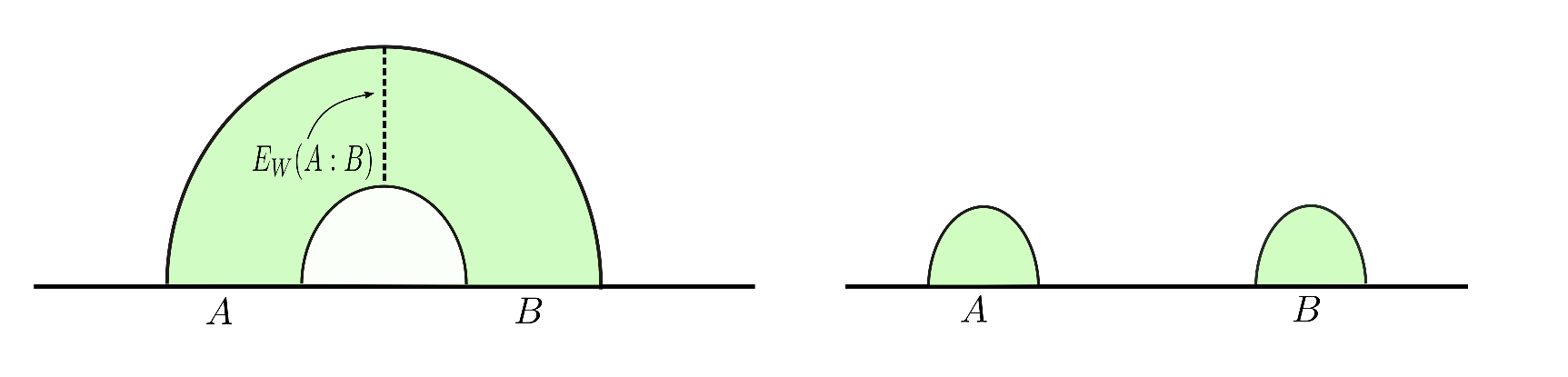}
\caption{Left: The colored region represents the entanglement wedge for subsystem $A\cup B$ in \poincare{} \ads{3}. The dotted line shows the entanglement wedge cross section. Right: If $A$ and $B$ are sufficiently far away, the entanglement wedge becomes disconnected and $E_W(A:B)=0$.}
\label{fg_wedge}
\end{figure}

We begin this section by reviewing the construction for the bulk entanglement wedge cross section (EWCS) in \ads{}/\cft{} described in \cite{Takayanagi:2017knl,Nguyen:2017yqw}. To this end it is required to consider spatial subsystems $A$ and $B$ in a \cft{d} dual to a static bulk \ads{d+1} geometry. Let $\Xi$ be a constant time slice in the bulk and $\Gamma_{AB}$ be an RT surface for the subsystem $A\cup B$. Then the codimension-one spatial region in $\Xi$ bounded by $A\cup B\cup\Gamma_{AB}$ is described as the entanglement wedge for the subsystems $A$ and $B$. Finally we consider the minimal area surface $\Sigma_{AB}^{\text{min}}$ which terminates on the boundary of the entanglement wedge, dividing the total entanglement wedge into two parts as illustrated in \cref{fg_wedge}.

The entanglement wedge cross section (EWCS), denoted by $E_W$, may then be defined as\footnote{Note that sometimes the minimal surface $\Sigma_{AB}^{\text{min}}$ itself is referred to as the entanglement wedge cross section. The meaning is usually clear from the context.}
\begin{equation}\label{static-wedge}
E_W(A:B)=\frac{\text{Area}(\Sigma_{AB}^{\text{min}})}{4G_N},
\end{equation}
where $G_N$ is the Newton constant. The reduced density matrix $\rho_{AB}$ is dual to the corresponding entanglement wedge \cite{Czech:2012bh,Wall:2012uf,Headrick:2014cta}. Note that the entanglement wedge includes the bulk region which is defined as the \textit{domain of dependence} for the spacelike homology surface $R_A$ bounded by the subsystem $A$ and its RT surface $\Gamma_A$. Some of the properties of the EWCS are listed below \cite{Takayanagi:2017knl,Nguyen:2017yqw}.
\begin{enumerate}
\item For a pure state $\rho_{AB}$, $E_W$ is equal to the entanglement entropy:
\begin{equation}\label{pure}
E_W(A:B)=S(A)=S(B).
\end{equation}
\item For a mixed state $\rho_{AB}$, $E_W$ is bounded above by the entanglement entropy:
\begin{equation}\label{mixed}
E_W(A:B)\leq\min[S(A),S(B)].
\end{equation}
\item For a mixed state $\rho_{AB}$, $E_W$ is bounded below by half the mutual information:
\begin{equation}
E_W(A:B)\geq\frac{I(A:B)}{2}.
\end{equation}
\item The $E_W$ is monotonic, \textit{i.e.}, it never increases upon discarding a subsystem:
\begin{equation}
E_W(A:BC)\geq E_W(A:B).
\end{equation}
\item For a tripartite system, $E_W$ has the following bound:
\begin{equation}
E_W(A:BC)\geq\frac{I(A:B)}{2}+\frac{I(A:C)}{2}.
\end{equation}
\item In a bipartite state that saturates the Araki-Lieb inequality, $S(AB)=\lvert S(A)-S(B)\rvert$, we have $E_W(A:B)=\min[S(A),S(B)]$.
\item For a tripartite pure state, the $E_W$ is polygamous:
\begin{equation}\label{tripartite}
E_W(A:BC)\leq E_W(A:B)+E_W(A:C).
\end{equation}
\end{enumerate}
Note that these properties conform to the properties of the entanglement of purification (EoP) described in quantum information theory \cite{2015PhRvA..91d2323B}.

\subsection{Computation of the EWCS}

We now proceed to review the computation of the EWCS in the \ads{3}/\cft{2} scenario. For this purpose it is required to consider a configuration of two subsystems $A$ and $B$ at zero temperature in the dual \cft{2} vacuum. These subsystems are described by spatial intervals $A=[a_1,a_2]$ and $B=[b_1,b_2]$ with lengths $l_1$ and $l_2$ respectively and are separated by distance $d$. The bulk dual for this case is a pure \ads{3} geometry in \poincare{} coordinates whose metric on a constant time slice is given by
\begin{equation}
ds^2=\frac{dx^2+dz^2}{z^2},
\end{equation}
where we have set the \ads{3} radius $R=1$. When the mutual information $I(A:B)>0$, the entanglement wedge remains connected otherwise it is disconnected. The EWCS for this configuration may be expressed as \cite{Takayanagi:2017knl}
\begin{equation}\label{min-cross-section}
E_W(A:B)=
\left\{
\begin{array}{ll}
\displaystyle
\frac{c}{6}\ln\left(1+2z+2\sqrt{z(z+1)}\right),&z>1,\\ \\
0,&0<z<1,
\end{array}
\right.
\end{equation}
where the conformal cross ratio $z$ is given by
\begin{equation}
z=\frac{(a_2-a_1)(b_2-b_1)}{(b_1-a_2)(b_2-a_1)}=\frac{l_1 l_2}{d(l_1+l_2+d)}.
\end{equation}
Note that \cref{min-cross-section} may be rewritten in the following form \cite{Kudler-Flam:2018qjo,Kudler-Flam:2019wtv}
\begin{equation}\label{ryu-ew}
E_W=
\left\{
\begin{array}{ll}
\displaystyle
\frac{c}{6}\ln\frac{1+\sqrt{x}}{1-\sqrt{x}},
&1/2<x<1,\\ \\
0,&0<x<1/2,
\end{array}
\right.
\end{equation}
where the cross ratio
\begin{equation}\label{cross_x}
x=\frac{(a_2-a_1)(b_2-b_1)}{(b_1-a_1)(b_2-a_2)}=\frac{l_1l_2}{(l_1+d)(l_2+d)}
\end{equation}
is related to $z$ through $z=x/(1-x)$.

For a finite temperature state in a \cft{2} defined on an infinite line, the bulk dual is a planar BTZ black hole (black string) whose metric is given as
\begin{equation}\label{static-btz}
ds^2=\frac{1}{z^2}\left(-f(z)dt^2+\frac{dz^2}{f(z)}+dx^2\right),\quad f(z)\equiv1-z^2/z_H^2,
\end{equation}
where the event horizon is located at $z=z_H$ and $z_H$ is related to the inverse temperature $\beta$ as $\beta=2\pi z_H$. The EWCS corresponding to the intervals $A=[a_1,a_2]$ and $B=[b_1,b_2]$ in a dual \cft{2} is given as follows \cite{Takayanagi:2017knl}
\begin{equation}\label{cross-section-temp}
E_W(A:B)= \frac{c}{6}\ln\left(1+2\zeta+2\sqrt{\zeta(\zeta+1)}\right),
\end{equation}
with $\zeta$ given as
\begin{equation}
\zeta\equiv\frac{\sinh\left(\frac{\pi(a_{2}-a_{1})}{\beta}\right)\sinh\left(\frac{\pi(b_{2}-b_{1})}{\beta}\right)}{\sinh\left(\frac{\pi(b_{1}-a_{2})}{\beta}\right)\sinh\left(\frac{\pi(b_{2}-a_{1})}{\beta}\right)}
=\frac{\sinh\left(\frac{\pi l_1}{\beta}\right)\sinh\left(\frac{\pi l_2}{\beta}\right)}{\sinh\left(\frac{\pi d}{\beta}\right)\sinh\left(\frac{\pi(l_1+l_2+d)}{\beta}\right)}.
\end{equation}

\subsubsection{EWCS for two disjoint intervals}\label{EWCSDJ}

For this configuration we will only describe the scenario where the two disjoint intervals are in proximity which corresponds to the regime $x\approx 1$ following \cite{Calabrese:2012nk,Kulaxizi:2014nma}. It was shown there that the entanglement negativity obtained through a replica technique in a \cft{2} was non universal, and a dominant universal form could be obtained only for the above proximity regime in the large central charge limit. Note that as explained in \cite{Calabrese:2012nk,Kulaxizi:2014nma} the regime $x\approx 1$ involves $l_1,l_2>>d$.\footnote{In the literature the regime $x\approx 1$ has been loosely stated as the limit $x\to 1$. However such a limit will force the EWCS to be divergent and implies setting $d\to 0$ which is not possible as $d>\epsilon$ where $\epsilon$ is the UV cutoff in the \cft{2}.} In this regime the EWCS for the two disjoint intervals in proximity is then obtained from \cref{ryu-ew} as follows
\begin{equation}\label{ew-disj-zero}
E_W(A:B)=\frac{c}{6}\ln\left(\frac{1}{1-x}\right)+\frac{c}{6}\ln 4.
\end{equation}

The EWCS for two disjoint intervals in proximity at zero temperature in a dual \cft{2} may be derived by substituting \cref{cross_x} into \cref{ew-disj-zero} as follows
\begin{equation}\label{ew-disj-zero-2}
E_W(A:B)=\frac{c}{6}\ln\left[\frac{(l_1+d)(l_2+d)}{d(l_1+l_2+d)}\right]+\frac{c}{6}\ln 4.
\end{equation}

For the intervals $A$ and $B$ in a finite size system of length $L$ at zero temperature, the EWCS may be obtained via the conformal map from the plane to cylinder $z\to w=(iL/2\pi)\ln z$ from \cref{ew-disj-zero} as
\begin{equation}\label{ew-disj-finite-size}
E_W(A:B)=\frac{c}{6}\ln\left[\frac{\sin\left(\frac{\pi(l_1+d)}{L}\right)\sin\left(\frac{\pi(l_2+d)}{L}\right)}{\sin\left(\frac{\pi d}{L}\right)\sin\left(\frac{\pi(l_1+l_2+d)}{L}\right)}\right]+\frac{c}{6}\ln 4.
\end{equation}

The EWCS for two disjoint intervals in proximity at a finite temperature may be obtained from \cref{ew-disj-zero} by employing the conformal transformation $z\to w=(\beta/2\pi)\ln z$ from the plane to the cylinder. It is then given by
\begin{equation}\label{ew-disj-finite-temp}
E_W(A:B)=\frac{c}{6}\ln\left[\frac{\sinh\left(\frac{\pi(l_1+d)}{\beta}\right)\sinh\left(\frac{\pi(l_2+d)}{\beta}\right)}{\sinh\left(\frac{\pi d}{\beta}\right)\sinh\left(\frac{\pi(l_1+l_2+d)}{\beta}\right)}\right]+\frac{c}{6}\ln 4.
\end{equation}

\subsubsection{EWCS for two adjacent intervals}

The mixed state configuration of two adjacent intervals may be constructed from the corresponding disjoint setup by taking the adjacent limit $d\to\epsilon$, where $\epsilon$ is the UV cutoff of the \cft{2} in the boundary. The EWCS for this configuration may now be obtained from \cref{ew-disj-zero-2} as follows
\begin{equation}\label{ew-adj}
E_W(A:B)=\frac{c}{6}\ln\left(\frac{l_1l_2}{\epsilon(l_1+l_2)}\right)+\frac{c}{6}\ln 4.
\end{equation} 

By taking the adjacent limit $d\to\epsilon$ in \cref{ew-disj-finite-size}, the EWCS for two adjacent intervals in a finite size system of length $L$ may expressed as
\begin{equation}\label{ew-adj-finite-size}
E_W(A:B)=\frac{c}{6}\ln\left[\left(\frac{L}{\pi\epsilon}\right)\frac{\sin\left(\frac{\pi l_1}{L}\right)\sin\left(\frac{\pi l_2}{L}\right)}{\sin\left(\frac{\pi(l_1+l_2)}{L}\right)}\right]+\frac{c}{6}\ln 4.
\end{equation}

The EWCS for two adjacent intervals at a finite temperature may be computed from \cref{ew-disj-finite-temp} by taking the adjacent limit $d\to\epsilon$ as follows
\begin{equation}\label{ew-adj-finite-temp}
E_W(A:B)=\frac{c}{6}\ln\left[\left(\frac{\beta}{\pi\epsilon}\right)\frac{\sinh\left(\frac{\pi l_1}{\beta}\right)\sinh\left(\frac{\pi l_2}{\beta}\right)}{\sinh\left(\frac{\pi(l_1+l_2)}{\beta}\right)}\right]+\frac{c}{6}\ln 4.
\end{equation}

\subsubsection{EWCS for a single interval}\label{sn_ew_sg}

The EWCS for a pure state configuration of a single interval $A$ of length $l$ at zero temperature may be obtained from the property of the EWCS for a pure state as described in \cref{pure}. It is given as
\begin{equation}\label{ew-single-zero}
E_W(A:B)=S(A)=\frac{c}{3}\ln\left(\frac{l}{\epsilon}\right).
\end{equation}

Similarly, the EWCS for a pure state configuration of a single interval $A$ in a finite size system of length $L$ may be computed using \cref{pure} and given by
\begin{equation}\label{ew-single-finite-size}
E_W(A:B)=\frac{c}{3}\ln\left(\frac{L}{\pi\epsilon}\sin\frac{\pi l}{L}\right).
\end{equation}

Subsequently, the authors in \cite{Takayanagi:2017knl} proposed the EWCS for a mixed state configuration of a single interval $A$ of length $l$ at finite temperature as the minimum of two possible candidates as expressed below
\begin{equation}
\label{ew_takayanagi_formula_1}
E_W(A:B)=\frac{c}{3}\min\left[\ln\left(\frac{\beta}{\pi\epsilon}\right),
~\ln\left(\frac{\beta}{\pi\epsilon}\sinh\frac{\pi l}{\beta}\right)\right].
\end{equation} 
The EWCS construction for this configuration will be described in detail in \cref{sn_hen_sg_tmp}.

\section{HEN from EWCS in \ads{3}/\cft{2}}\label{sn_HEN_EWCS}

In this section we describe the construction in \cite{Kudler-Flam:2018qjo,Kusuki:2019zsp} which proposed the entanglement wedge cross section as the holographic dual of the entanglement negativity. The authors in \cite{Takayanagi:2017knl,Nguyen:2017yqw} had demonstrated that the EWCS was dual to the entanglement of purification (EoP) and followed all the properties of the EoP in quantum in formation theory \cite{2015PhRvA..91d2323B}. Motivated by these developments and results from quantum error correcting codes, the authors in \cite{Kudler-Flam:2018qjo,Kusuki:2019zsp} conjectured that the EWCS backreacted by a bulk cosmic brane for the conical defect geometry is also dual to the entanglement negativity for configurations involving spherical entangling surfaces. The holographic entanglement negativity for the corresponding dual \cft{}s was then expressed in terms of the bulk EWCS as follows \cite{Kudler-Flam:2018qjo,Kusuki:2019zsp} 
\begin{equation}\label{holo-neg}
\mathcal{E}=\mathcal{X}_dE_W,
\end{equation}
where $\mathcal{X}_d$ is the same dimension dependent constant described in \cref{HREE}. The above conjecture was substantiated more concisely for holographic \cft{}s as \cite{Kusuki:2019zsp}
\begin{equation}\label{holo-neg-sr}
\mathcal{E}=\frac{S_R^{(1/2)}}{2},
\end{equation}
where $S_R$ is a correlation measure termed as the reflected entropy, and $S_R^{(1/2)}$ is the \renyi{} reflected entropy of order half. In the next few subsections we briefly review the computation of the holographic entanglement negativity from the EWCS for various bipartite states described by two disjoint intervals, two adjacent intervals, and a single interval in the context of the \ads{3}/\cft{2} scenario.

\subsection{Negativity for two disjoint intervals}

\subsubsection{Zero temperature}

The computation of the entanglement negativity for two disjoint intervals involves the four point twist correlator whose explicit form contains an arbitrary non universal function of the cross ratio and depends on the full operator content of the corresponding \cft{2}. Using Zamolodchikov recursion relations, the authors in \cite{Kusuki:2019zsp} have numerically shown that the entanglement negativity for disjoint intervals is proportional to the EWCS. However a recent development indicates the presence of an extra additive constant in the EWCS for two disjoint intervals in a \cft{2}, which may be determined through a careful bulk computation \cite{Hayden:2021gno}. As described earlier in \cref{EWCSDJ} the entanglement negativity for two disjoint intervals in a \cft{2} is non universal for general values of the cross ratio $x$ and a universal divergent behavior in the large central charge limit may be obtained for the regime $x\approx 1$ where the intervals are in proximity ($l_1,l_2>>d$).\footnote{A brief review of the determination of the entanglement negativity for disjoint intervals from a field theory replica technique approach as described in \cite{Calabrese:2012nk,Kulaxizi:2014nma} has been provided in \cref{app_hendj}.} For this regime on using \cref{holo-neg,ew-disj-zero}, the holographic entanglement negativity for the mixed state of two disjoint intervals in proximity at zero temperature may be expressed as
\begin{equation}\label{neg-disj-zero}
\mathcal{E}=\frac{c}{4}\ln\left[\frac{(l_1+d)(l_2+d)}{d(l_1+l_2+d)}\right]+\frac{c}{4}\ln 4,
\end{equation}
where $l_1$, $l_2$ are lengths of the intervals, and $d$ is the length of the separation between the intervals $A$ and $B$. Note that the above result is cutoff independent. The above result matches with the corresponding \cft{2} replica result \cite{Calabrese:2012ew,Calabrese:2012nk, Kulaxizi:2014nma} obtained by monodromy technique for the $t$ channel in the regime $x\approx 1$ modulo the second constant term. As discussed in the \nameref{sec_intro}, the constant second term in \cref{neg-disj-zero} arises from the Markov gap.\footnote{\label{fn_Markov_dj}The holographic Markov gap between the reflected entropy and the mutual information is briefly described in \cref{app_gap}. For details see \cite{Hayden:2021gno}.} However the result of holographic entanglement negativity for the two disjoint intervals obtained in \cite{Malvimat:2018txq} matches exactly with the corresponding field theory replica technique results. This illustrates the equivalence of the two holographic proposals for the entanglement negativity expressed in \cref{HENDJ,holo-neg} up to an additive constant.

\subsubsection{Finite size}

The holographic entanglement negativity for the configuration of two disjoint intervals in a finite size \cft{2} may be obtained using \cref{holo-neg,ew-disj-finite-size} as
\begin{equation}\label{neg-disj-finite-size}
\mathcal{E}=\frac{c}{4}\ln\left[\frac{\sin\left(\frac{\pi(l_1+d)}{L}\right)\sin\left(\frac{\pi(l_2+d)}{L}\right)}{\sin\left(\frac{\pi d}{L}\right)\sin\left(\frac{\pi(l_1+l_2+d)}{L}\right)}\right]+\frac{c}{4}\ln 4.
\end{equation}
Note again that the result in \cref{neg-disj-finite-size} is cutoff independent. We observe again that apart from the constant term, the holographic entanglement negativity for two disjoint intervals in \cref{neg-disj-finite-size} matches with the corresponding result from an earlier alternative proposal described in \cref{HENDJ} \cite{Malvimat:2018txq} demonstrating their equivalence.

\subsubsection{Finite temperature}

For the mixed state configuration of two disjoint intervals at a finite temperature, the holographic entanglement negativity may be computed using \cref{holo-neg,ew-disj-finite-temp} as
\begin{equation}\label{neg-disj-finite-temp}
\mathcal{E}=\frac{c}{4}\ln\left[\frac{\sinh\left(\frac{\pi(l_1+d)}{\beta}\right)\sinh\left(\frac{\pi(l_2+d)}{\beta}\right)}{\sinh\left(\frac{\pi d}{\beta}\right)\sinh\left(\frac{\pi(l_1+l_2+d)}{\beta}\right)}\right]+\frac{c}{4}\ln 4.
\end{equation}
Note that the expression in \cref{neg-disj-finite-temp} is once again cutoff independent. The above equation matches with the entanglement negativity obtained by the field theory replica technique in \cft{2}s \cite{Calabrese:2012nk}, modulo the additive constant in the large central charge limit. Once again we observe that the holographic entanglement negativity in \cref{neg-disj-finite-temp} matches with that obtained from the alternative proposal described in \cref{HENDJ} \cite{Malvimat:2018txq} modulo the constant term, establishing their equivalence.

\subsection{Negativity for two adjacent intervals}

\subsubsection{Zero temperature}

The holographic entanglement negativity for the bipartite mixed state configuration described by two adjacent intervals at zero temperature in a \cft{2} may be obtained from the backreacted EWCS using \cref{holo-neg,ew-adj} as follows
\begin{equation}\label{adj-zero}
\mathcal{E}=\frac{c}{4}\ln\left(\frac{l_1l_2}{\epsilon(l_1+l_2)}\right)+\frac{c}{4}\ln 4,
\end{equation}
where $\epsilon$ is the UV cutoff. We observe that the above result matches with the entanglement negativity obtained by the field theory replica technique in the large central charge limit \cite{Calabrese:2012nk} modulo the constant term related to the holographic Markov gap (see \cref{fn_Markov_dj}). Note however that the holographic entanglement negativity for this mixed state configuration obtained through the alternate conjecture involving an algebraic sum of lengths of bulk geodesics in \cite{Jain:2017aqk} matches exactly with the field theory results in the large $c$ limit. This demonstrates the equivalence of the two holographic proposals for the entanglement negativity modulo the constant from the Markov gap.

\subsubsection{Finite size}

The holographic entanglement negativity for the configuration of two adjacent intervals for a finite size \cft{2} may be obtained using \cref{holo-neg,ew-adj-finite-size} as
\begin{equation}\label{neg-adj-finite-size}
\mathcal{E}=\frac{c}{4}\ln\left[\left(\frac{L}{\pi\epsilon}\right)\frac{\sin\left(\frac{\pi l_1}{L}\right)\sin\left(\frac{\pi l_2}{L}\right)}{\sin\left(\frac{\pi(l_1+l_2)}{L}\right)}\right]+\frac{c}{4}\ln 4,
\end{equation}
where $\epsilon$ is the UV cutoff. The above result also matches with the corresponding replica technique results \cite{Calabrese:2012nk} in the large central charge limit up to the Markov gap constant. Note that the above result in \cref{neg-adj-finite-size} modulo the constant may also be obtained using the alternative proposal described in \cref{HENADJ} \cite{Jain:2017aqk} which involves an algebraic sum of lengths of the bulk geodesics homologous to appropriate combinations of the intervals and illustrates their equivalence.

\subsubsection{Finite temperature}

The EWCS for two adjacent intervals at a finite temperature may be computed from \cref{cross-section-temp} by taking the adjacent limit $d\to\epsilon$. Now using \cref{holo-neg}, the holographic entanglement negativity for the finite temperature mixed state configuration of two adjacent intervals in a \cft{2} may be expressed as
\begin{equation}\label{adj-finite}
\mathcal{E}=\frac{c}{4}\ln\left[\left(\frac{\beta}{\pi\epsilon}\right)\frac{\sinh\left(\frac{\pi l_1}{\beta}\right)\sinh\left(\frac{\pi l_2}{\beta}\right)}{\sinh\left(\frac{\pi(l_1+l_2)}{\beta}\right)}\right]+\frac{c}{4}\ln 4,
\end{equation}
where $\epsilon$ is the UV cutoff. As discussed earlier, the constant second term in the above equation may be related to the holographic Markov gap. The first term in \cref{adj-finite} matches with the corresponding result for holographic entanglement negativity in \cite{Jain:2017aqk} using the alternate construction given in \cref{HENADJ} and the \cft{2} replica technique result \cite{Calabrese:2012nk} in the large central charge limit. This once again demonstrates the equivalence of the two holographic entanglement negativity proposals modulo the Markov gap constant.

\subsection{Negativity for a single interval}\label{sn_hen_sg}

In this section, we consider the bipartite configuration of a single interval in a dual \cft{2}. To this end it is required to consider the system $A\cup B$ which includes the interval $A$ of length $l$ and rest of the system is denoted as $B$.

\subsubsection{Zero temperature}

The holographic entanglement negativity for the pure state of a single interval $A$ in a \cft{2} at zero temperature may be obtained by considering the property of the EWCS described in \cref{pure}. Utilizing the holographic prescription from \cref{holo-neg} and expression for the corresponding EWCS in \cref{ew-single-zero}, the entanglement negativity for this pure state is given as
\begin{equation}\label{sin-zero}
\mathcal{E}=\frac{c}{2}\ln\left(\frac{l}{\epsilon}\right).
\end{equation}
The above result matches exactly with the corresponding \cft{2} replica technique result as described in \cite{Calabrese:2012ew,Calabrese:2012nk} in the large central charge limit. Interestingly, here we observe the absence of any contribution related to the holographic Markov gap as the configuration of a single interval in zero temperature \cft{2} is in a pure state. As earlier the holographic entanglement negativity in \cref{sin-zero} matches with that obtained from the alternative proposal in \cref{HENSIN} \cite{Chaturvedi:2016rcn} once again illustrating their equivalence.

\subsubsection{Finite size}

The holographic entanglement negativity for the pure state configuration of a single interval in a finite size system may be computed using \cref{ew-single-finite-size,holo-neg} as follows
\begin{equation}\label{en-single-finite-size}
\mathcal{E}=\frac{c}{3}\ln\left(\frac{L}{\pi\epsilon}\sin\frac{\pi l}{L}\right).
\end{equation}
Note that the above result may also be obtained using the alternative holographic proposal as described in \cref{HENSIN}.

\subsubsection{Finite temperature}\label{sn_hen_sg_tmp}

\begin{figure}
\centering
\includegraphics[scale=.45]{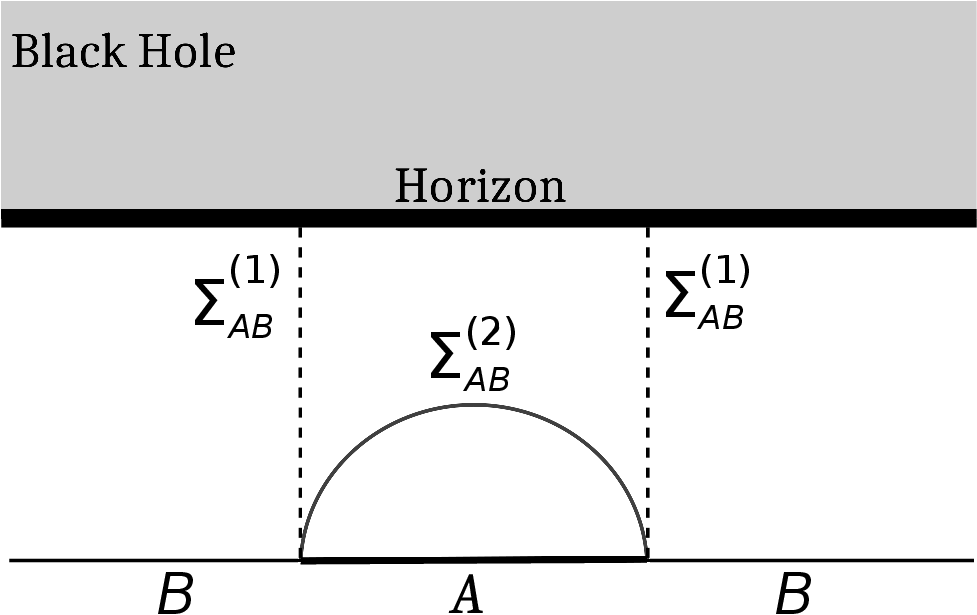}
\caption{EWCS for a single interval $A$ (with rest of the system $B$) in a thermal \cft{2} on the boundary, dual to a bulk planar BTZ black hole geometry.}
\label{Single-interval}
\end{figure}

We begin with a brief review of the entanglement wedge construction in the context of a single interval in a \cft{2} at a finite temperature, dual to a bulk planar BTZ black hole as described in \cite{Takayanagi:2017knl}. The authors in \cite{Takayanagi:2017knl} considered the bipartition (\cref{Single-interval}) comprising a single interval (of length $l$), denoted by $A$, with the rest of the system denoted by $B$. Note that in \cite{Takayanagi:2017knl}, $\Sigma_{AB}^{\text{min}}$, the minimal cross section of the entanglement wedge for the intervals $A$ and $B$, has two possible candidates. The first one, denoted by $\Sigma_{AB}^{(1)}$, is the union of the two dotted lines depicted in \cref{Single-interval}, while the other one, $\Sigma_{AB}^{(2)}$, is given by the RT surface $\Gamma_A$. Then, the EWCS was given as \cite{Takayanagi:2017knl}
\begin{align}
E_W(A:B)&=\frac{c}{3}\min\left[\text{Area}\left(\Sigma_{AB}^{(1)}\right),
\text{Area}\left(\Sigma_{AB}^{(2)}\right)\right]\label{ew_takayanagi_formula_2}\\
&=\frac{c}{3}\min\left[\ln\left(\frac{\beta}{\pi\epsilon}\right),
~\ln\left(\frac{\beta}{\pi\epsilon}\sinh\frac{\pi l}{\beta}\right)\right],\label{ew_takayanagi_result}
\end{align}
where $\epsilon$ is the UV cutoff, and $\beta$ is the inverse temperature.
Using the result of \cref{ew_takayanagi_result} the authors in \cite{Kudler-Flam:2018qjo,Kusuki:2019zsp} described the holographic entanglement negativity of the single interval in a \cft{2} at a finite temperature as
\begin{equation}\label{naive}
\mathcal{E}=\frac{c}{2}\min\left[\ln\left(\frac{\beta}{\pi\epsilon}\sinh\frac{\pi l}{\beta}\right), ~\ln\left(\frac{\beta}{\pi\epsilon}\right)\right].
\end{equation}
The authors substantiated their results utilizing the monodromy technique to extract the dominant contribution in the $s$ and $t$ channels for the relevant four point twist correlator. However the above results do not exactly reproduce the corresponding replica technique results for the mixed state configuration in question described in \cite{Calabrese:2014yza} except in the low and high temperature limits. Specifically the subtracted thermal entropy term is missing from the holographic entanglement negativity and this issue requires further analysis. In \cref{sn_issues} we will carefully investigate the above issue for the computation of the holographic entanglement negativity from the EWCS in \cite{Kudler-Flam:2018qjo,Kusuki:2019zsp} for the configuration in question.

\section{Issue with the thermal entropy term}\label{sn_issues}

Calabrese, Cardy and Tonni in a significant communication \cite{Calabrese:2014yza} computed the entanglement negativity of a single interval (of length $l$) for a \cft{2} at a finite temperature, which is given by
\begin{equation}\label{en_sg_cardy}
\mathcal{E}=\frac{c}{2}\ln\left(\frac{\beta}{\pi\epsilon}\sinh\frac{\pi l}{\beta}\right)-\frac{\pi cl}{2\beta} +f\left(e^{-2\pi l/\beta}\right)+2\ln c_{1/2},
\end{equation}
where $\epsilon$ is the UV cutoff. Here $f$ is an arbitrary function and $c_{1/2}$ is a constant, which are non universal and depend on the full operator content of the theory.

Comparing \cref{naive} with \cref{en_sg_cardy}, we note that the holographic entanglement negativity, as computed from \cref{holo-neg}, does not match exactly\footnote{In \cite{Takayanagi:2017knl}, the authors have indicated that for a single interval at a finite temperature with a length $l\gg\beta\ln(\sqrt{2}+1)/\pi$, the extensive contribution is missing in the expression for the EWCS, as described in \cref{ew_takayanagi_result}.} with the corresponding replica technique result reported in \cite{Calabrese:2014yza}, in the large central charge limit. Specifically the subtracted thermal entropy term in the large $c$ replica technique result is missing in the expression for the holographic entanglement negativity of a single interval at a finite temperature described in \cite{Kudler-Flam:2018qjo,Kusuki:2019zsp}. One may further observe that the entanglement negativity in \cref{en_sg_cardy} reduces to that in \cref{naive} only in the specific limits of low temperature ($\beta\to\infty$) and high temperature ($\beta\to 0$). In the next subsection we briefly describe the monodromy analysis of the four point twist correlator relevant for this configuration towards a resolution of the above issue with the subtracted thermal entropy term for the entanglement negativity.

\subsection{Large central charge limit}\label{subsec_monodromy}

We begin by reviewing the results obtained through the monodromy technique employed in \cite{Kusuki:2019zsp} to compute the entanglement negativity of a single interval as given in \cref{naive}. To this end it is required to consider two auxiliary intervals $B_1=[-L,0]$ and $B_2=[l,L]$ on either side of the single interval $A=[0,l]$ (see \cite{Calabrese:2014yza} for details). The rest of the system is denoted by $B\equiv B_1\cup B_2$. Finally we implement the bipartite limit $B\to A^c$ through $L\to\infty$ to arrive at the required configuration for the single interval $A$ and the rest of the system $B=A^c$. The entanglement negativity of a single interval at zero temperature in a \cft{2} may then be described by the following four point twist correlator on the complex plane \cite{Calabrese:2014yza}
\begin{equation}\label{en_4pt}
\mathcal{E}=\lim_{L\to\infty}\lim_{n_e\to 1}\ln\left\langle\mathcal{T}_{n_e}(-L)
\overline{\mathcal{T}}_{n_e}^2(0)\overline{\mathcal{T}}_{n_e}^2(l)\mathcal{T}_{n_e}(L)\right\rangle_\mathbb{C}.
\end{equation}
Through a suitable conformal transformation (see \cite{Malvimat:2018txq} for a detailed review), \cref{en_4pt} may be recast as \cite{Kusuki:2019zsp}
\begin{equation}\label{en_4pt_x}
\mathcal{E}=\lim_{L\to\infty}\lim_{n_e\to 1}\ln\left\langle\mathcal{T}_{n_e}(0)
\overline{\mathcal{T}}_{n_e}^2(x)\overline{\mathcal{T}}_{n_e}^2(1)\mathcal{T}_{n_e}(\infty)\right\rangle_\mathbb{C}.
\end{equation}

For a \cft{2} at a finite temperature $1/\beta$, the entanglement negativity for a single interval may be computed from \cref{en_4pt} or \cref{en_4pt_x} through the conformal transformation $z\to w=(\beta/2\pi)\ln z$ from the complex plane to the cylinder. The entanglement negativity for a single interval in a thermal \cft{2} may then be obtained as follows \cite{Calabrese:2014yza,Kusuki:2019zsp}
\begin{align}
\mathcal{E}&=\lim_{L\to\infty}\lim_{n_e\to 1}\ln\left\langle\mathcal{T}_{n_e}(-L)\overline{\mathcal{T}}_{n_e}^2(0)
\overline{\mathcal{T}}_{n_e}^2(l)\mathcal{T}_{n_e}(L)\right\rangle_{\text{cyl}(\beta)}\nonumber\\
&=\lim_{L\to\infty}\lim_{n_e\to 1}\ln\left\langle\mathcal{T}_{n_e}(0)
\overline{\mathcal{T}}_{n_e}^2(x)\overline{\mathcal{T}}_{n_e}^2(1)\mathcal{T}_{n_e}(\infty)\right\rangle_\mathbb{C}
+\frac{c}{2}\ln\left(\frac{\beta}{2\pi\epsilon}e^{\frac{\pi l}{\beta}}\right),\label{en_4pt_ft}
\end{align}
where the cross ratio $x$ is specified by $\lim\limits_{L\to\infty}x=e^{-2\pi l/\beta}$.

In the large central charge limit, the four point twist correlator in \cref{en_4pt_ft} may be expressed in terms of a dominant single conformal block in the $s$ and $t$ channels depending on the cross ratio $x$ as described in \cite{Malvimat:2017yaj}.

For the $s$ channel (described by $x\approx 0$), the authors in \cite{Kusuki:2019zsp} have computed the four point function on the complex plane as
\begin{equation}\label{4pt_s_ch}
\lim_{L\to\infty}\lim_{n_e\to 1}\ln\left\langle\mathcal{T}_{n_e}(0)\overline{\mathcal{T}}_{n_e}^2(x)
\overline{\mathcal{T}}_{n_e}^2(1)\mathcal{T}_{n_e}(\infty)\right\rangle_\mathbb{C}=\frac{c}{4}\ln x.
\end{equation}
Utilizing the above four point twist correlator, the entanglement negativity may be computed from \cref{en_4pt_ft} as\footnote{Note that we have omitted a Markov gap term $-(c/4)\ln 4$ on the right hand side of \cref{en_s_ch}.}
\begin{equation}\label{en_s_ch}
\mathcal{E}=\frac{c}{2}\ln\left(\frac{\beta}{\pi\epsilon}\right).
\end{equation}

For the $t$ channel (given by $x\approx 1$), the authors in \cite{Kusuki:2019zsp} have obtained the following four point function
\begin{equation}\label{4pt_t_ch}
\lim_{L\to\infty}\lim_{n_e\to 1}\ln\left\langle\mathcal{T}_{n_e}(0)\overline{\mathcal{T}}_{n_e}^2(x)
\overline{\mathcal{T}}_{n_e}^2(1)\mathcal{T}_{n_e}(\infty)\right\rangle_\mathbb{C}=\frac{c}{2}\ln\left(1-x\right),
\end{equation}
utilizing which the entanglement negativity may be obtained from \cref{en_4pt_ft} as
\begin{equation}\label{en_t_channel}
\mathcal{E}=\frac{c}{2}\ln\left(\frac{\beta}{\pi\epsilon}\sinh\frac{\pi l}{\beta}\right).
\end{equation}

We note that the four point functions on the complex plane given in \cref{4pt_s_ch,4pt_t_ch}, utilized to compute the entanglement negativity for both the channels match with those obtained in \cite{Malvimat:2017yaj}. We further observe that the monodromy technique employed in \cite{Chaturvedi:2016rcn,Malvimat:2017yaj}, and the monodromy method utilized by the authors in \cite{Kusuki:2019zsp}, all produce identical large central charge limit for the four point function.

However we would like to emphasize here that the authors in \cite{Kusuki:2019zsp}, motivated by the large $c$ computations for the entanglement entropy in \cite{Hartman:2013mia, Headrick:2010zt}, assumed that the $s$ and $t$ channel results are valid beyond their usual regimes $x\approx 0$ and $x\approx 1$, that is, for $0<x<1/2$ and $1/2<x<1$ respectively. In other words their computation involves an assumption that there is a phase transition for the large central charge limit for the entanglement negativity of a single interval at $x=1/2$. Although this is true for the entanglement entropy \cite{Hartman:2013mia, Headrick:2010zt}, for this specific case of the entanglement negativity for a single interval, this assumption is not valid as the required four point twist correlator is obtained from a specific channel for the corresponding six point correlator as described in \cite{Malvimat:2017yaj}. The above conclusion is also supported by an alternate EWCS construction for this configuration proposed in \cref{sec_alt_construct} to resolve the issue with the missing thermal term for the holographic entanglement negativity in \cite{Kudler-Flam:2018qjo,Kusuki:2019zsp}.

\subsection{Alternate EWCS construction}\label{sec_alt_construct}

In this subsection we propose an alternative construction to that described in \cite{Takayanagi:2017knl} for the mixed state configuration of a single interval at a finite temperature in \ads{3}/\cft{2}. To this end we consider the following properties [refer to \cref{tripartite}] of the EWCS for tripartite pure state configurations comprising subsystems $A$, $B$ and $C$ (see \cite{Takayanagi:2017knl,Nguyen:2017yqw} for a review)
\begin{equation}\label{tripartite_ub}
E_W(A:BC)\leq E_W(A:B)+E_W(A:C),
\end{equation}
\begin{equation}\label{tripartite_lb}
\frac{1}{2}I(A:B)+\frac{1}{2}I(A:C)\leq E_W(A:BC),
\end{equation}
where $I(A:B)$ is the mutual information between $A$ and $B$. For two adjacent intervals $A$ and $B$ at a finite temperature in a \cft{2} dual to a bulk planar BTZ black hole, the EWCS may be explicitly computed through the adjacent limit in the corresponding disjoint interval result derived in \cite{Takayanagi:2017knl}. The holographic mutual information for these adjacent intervals may also be explicitly calculated from the corresponding entanglement entropies. Comparing these results, we obtain the following relation for this specific configuration
\begin{equation}\label{tripartite_ewcs_mi}
E_W(A:B)=\frac{1}{2}I(A:B).
\end{equation}
Substituting the result described in \cref{tripartite_ewcs_mi} into \cref{tripartite_lb}, and comparing with \cref{tripartite_ub}, we arrive at the following equality for the bulk BTZ black hole configuration
\begin{equation}\label{tripartite_equality}
E_W(A:BC)=E_W(A:B)+E_W(A:C),
\end{equation}
where $B$ and $C$ are adjacent to $A$.

\begin{figure}
\centering
\includegraphics[scale=.45]{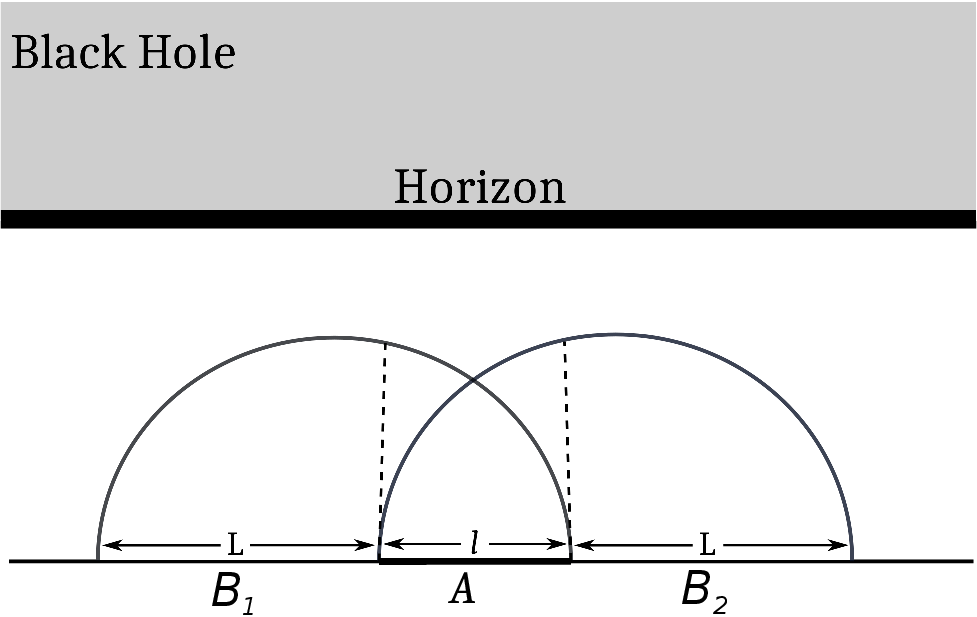}
\caption{Alternate computation of the EWCS (dotted lines) for a single interval $A$ (with rest of the system $B$) in a finite temperature \cft{2}, dual to a planar bulk BTZ black hole geometry.}
\label{Single-interval-modified}
\end{figure}

Following the construction in \cite{Calabrese:2014yza}, we now consider a tripartition (see \cref{Single-interval-modified}) consisting of interval $A$ of length $l$, with two auxiliary intervals $B_1$ and $B_2$, each of length $L$, on either side of $A$ where $B\equiv B_1\cup B_2$. Next we implement the bipartite limit $L\to\infty$ to recover the original configuration with a single interval $A$ and the rest of the system given by $B=A^c$. Note that in the bipartite limit, $A\cup B$ describes the full system which is in a pure state and obeys \cref{tripartite_equality}. Thus for this configuration (in the bipartite limit) 
\begin{equation}\label{tripartite_equality_specific}
\lim_{L\to\infty}E_W(A:B_1 B_2)=\left.\lim_{L\to\infty}\middle[E_W(A:B_1)+E_W(A:B_2)\right].
\end{equation}
Computing the right hand side of \cref{tripartite_equality_specific}, we obtain the EWCS for the bipartition involving the interval $A$ and rest of the system $B =A^c$ as follows
\begin{equation}\label{ew_single_correct}
E_W(A:B)=\frac{c}{3}\ln\left(\frac{\beta}{\pi\epsilon}\sinh\frac{\pi l}{\beta}\right)-\frac{\pi cl}{3\beta}+\frac{c}{3}\ln 4.
\end{equation}
The holographic entanglement negativity for the mixed state configuration of a single interval $A$ in a \cft{2} at a finite temperature dual to the bulk planar BTZ black hole may now be obtained by utilizing \cref{holo-neg,ew_single_correct} as follows
\begin{equation}\label{en_single_correct}
\mathcal{E}=\frac{c}{2}\ln\left(\frac{\beta}{\pi\epsilon}\sinh\frac{\pi l}{\beta}\right)-\frac{\pi cl}{2\beta}+\frac{c}{2}\ln 4.
\end{equation}
The above expression matches exactly with the corresponding field theory replica technique result for the entanglement negativity described in \cref{en_sg_cardy} in the large central charge limit up to a constant related to the Markov gap. Interestingly, the result for the holographic entanglement negativity for this mixed state configuration utilizing the alternate holographic construction involving the algebraic sum of bulk geodesics as reported in \cite{Chaturvedi:2016rcn} matches exactly with the corresponding field theory replica technique results in the large $c$ limit without the constant. This once again indicates the equivalence of the two proposals up to a constant.

\begin{figure}
\centering
\begin{subfigure}{0.8\textwidth}
\centering
\includegraphics[width=0.95\linewidth]{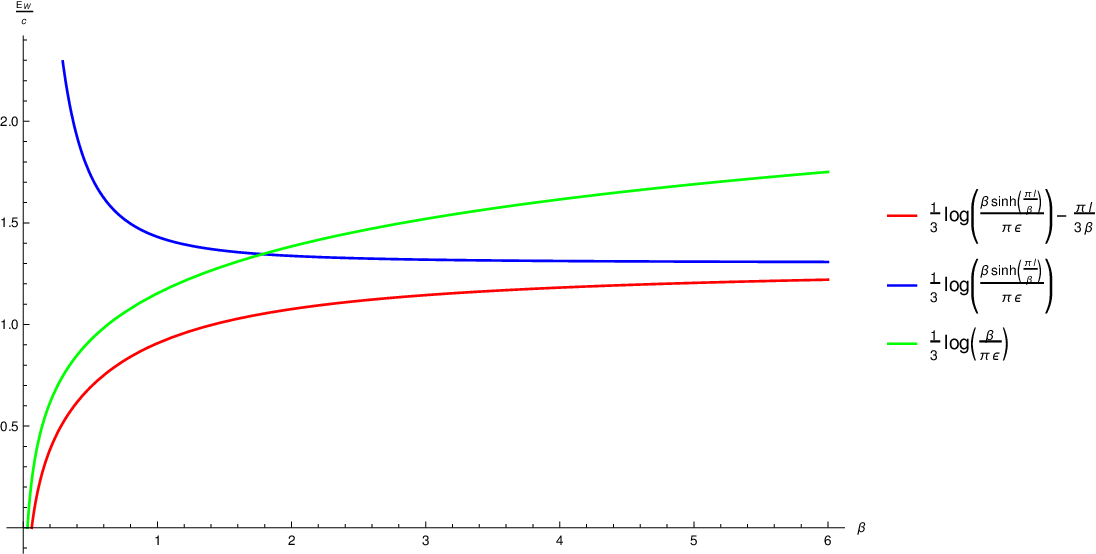}
\caption{$E_W/c$ vs.\@ $\beta$ plots for various candidates for EWCS.}
\label{plot1}
\end{subfigure}
\\ 
\bigskip
\begin{subfigure}{.8\textwidth}
\centering
\includegraphics[width=0.95\linewidth]{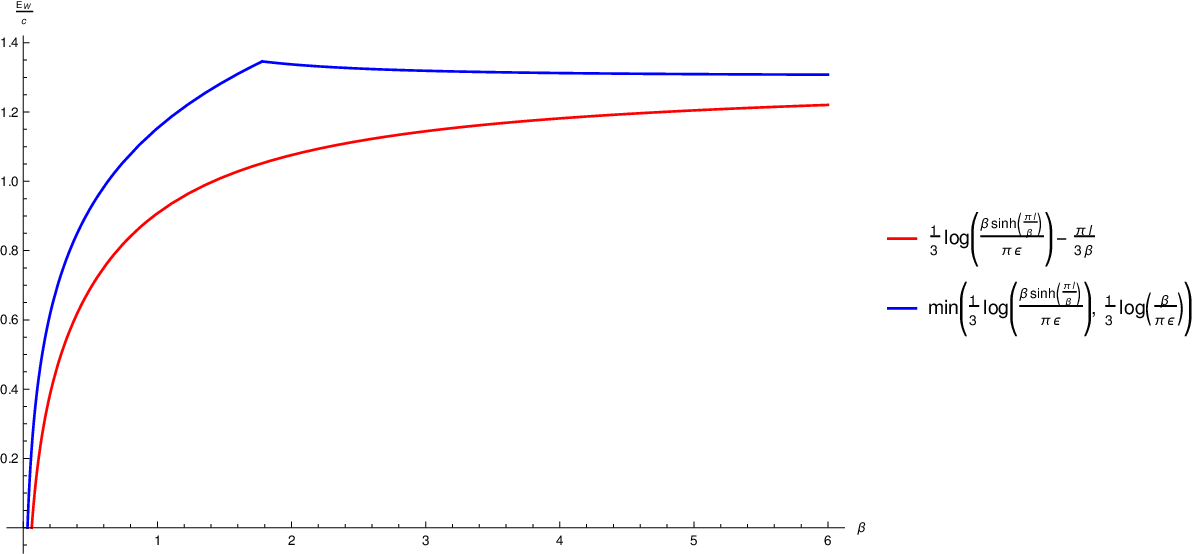}
\caption{$E_W/c$ vs.\@ $\beta$ plots for EWCS utilizing the two different constructions.}
\label{plot2}
\end{subfigure}
\caption{Plots for different choices of EWCS against the inverse temperature for a single interval at a finite temperature. Here, $\epsilon=0.01$ and $l=0.5$.}
\label{plot}
\end{figure}

In \cref{plot} we have plotted the possible candidates for the EWCS as a function of the inverse temperature $\beta$ to compare our construction with that described in \cite{Takayanagi:2017knl} for this configuration. In \cref{plot1}, the two possible candidates (green and blue curves) for the EWCS described in \cref{ew_takayanagi_result} have been compared with our expression given in \cref{ew_single_correct} without the third term (red curve). In \cref{plot2}, the proposed EWCS (blue curve) as prescribed in \cite{Takayanagi:2017knl}, given in \cref{ew_takayanagi_result}, has been plotted along with the EWCS modulo the constant (red curve) obtained by utilizing our alternative proposal as described in \cref{ew_single_correct}. It is interesting to note that our proposed EWCS always remains strictly less than that advanced in \cite{Takayanagi:2017knl} for the range of $\beta$ used in the plot. This conclusively singles out our construction for the EWCS over the other proposal as the correct minimal EWCS, which also reproduces the replica technique result described in \cite{Calabrese:2014yza}, in the large $c$ limit. We also observe that \cref{ew_single_correct,ew_takayanagi_result} asymptotically approach each other in the limits of high ($\beta\to 0$) and low ($\beta\to\infty$) temperatures. Naturally this resolves the issue with the holographic entanglement negativity for this configuration and restores the missing thermal entropy term.

\section{Discussions and conclusions}\label{sec_summary}

To summarize we have established the equivalence modulo the Markov gap constant of two different proposals in the literature for the holographic entanglement negativity of bipartite states in the context of the \ads{3}/\cft{2} correspondence. The first proposal described in \cite{Chaturvedi:2016rcn,Jain:2017aqk,Malvimat:2018txq} involved an algebraic sum of the lengths of bulk geodesics homologous to certain combinations of subsystems, and the second one reported in \cite{Kudler-Flam:2018qjo,Kusuki:2019zsp} was based on the EWCS backreacted by a cosmic brane for the conical defect of the replicated bulk geometry in a gravitational path integral.

In this connection we have analyzed and compared the results for the holographic entanglement negativity following from the above two proposals for various bipartite state configurations described by two disjoint intervals, two adjacent intervals, and a single interval in dual \cft{2}s. We observe that the results obtained from these two proposals match with each other up to certain constants which arise from the Markov gap for the EWCS, establishing their equivalence.

In this context, we have investigated a critical issue with the proposal involving the backreacted EWCS for the holographic entanglement negativity of a single interval in a finite temperature \cft{2} dual to a bulk planar BTZ black hole. Specifically the holographic entanglement negativity obtained for the above configuration from the backreacted EWCS excluded a subtracted thermal entropy term in the corresponding field theory replica technique result at large $c$. In this article, we have resolved this significant issue by proposing an alternative construction for the EWCS of this configuration which is actually the minimal one. Our construction involved the introduction of two large auxiliary intervals adjacent to the single interval in question. Subsequently a bipartite limit for this configuration was implemented through the consideration of the auxiliary intervals to be infinite and constitute the rest of the system. To this end we utilized certain polygamy properties to obtain the correct bulk EWCS for the single interval in a holographic \cft{2}. Interestingly our results following from the above construction reproduced the excluded thermal term and matched exactly with the field theory replica technique result in the large central charge limit in the literature. Naturally the holographic entanglement negativity proposal involving the backreacted EWCS requires substantiation from explicit higher dimensional examples for generic \ads{d+1}/\cft{d} scenarios. However the difficulties for the computation of the EWCS for higher dimensions are well known. We hope to address some of these issues in the near future.

\section*{Acknowledgments}

We acknowledge Jonah Kudler-Flam, Yuya Kusuki and Shinsei Ryu for communication, which led to this work. We would like to thank Saikat Ghosh, Debarshi Basu and Vinayak Raj for crucial discussions. The work of JKB is supported by the National Science and Technology Council of Taiwan with the grant 112-2636-M-110-006. The work of VM is supported by the NRF grant funded by the Korea government (MSIT) (No. 2022R1A2C1003182) and by the Brain Pool program funded by the Ministry of Science and ICT through the National Research Foundation of Korea (RS-2023-00261799). The work of HP is supported by the NCTS, Taiwan. The work of GS is partially supported by the Dr.\@ Jag Mohan Garg Chair Professor position at the Indian Institute of Technology, Kanpur.

\appendix

\section{Entanglement negativity of two disjoint intervals}\label{app_hendj}

In this appendix following \cite{Calabrese:2012ew,Calabrese:2012nk,Kulaxizi:2014nma} we describe the derivation of the entanglement negativity for two disjoint intervals in a \cft{2} through a replica technique in the large central charge limit. The entanglement negativity for this case is described by a four point twist field correlator involving a non universal function of the cross ratio $x$, which depends on the full operator content of the theory. As described in \cite{Calabrese:2012nk} it is not possible to evaluate this non universal function analytically for general values of the cross ratio $x$. However, it may be approximated in the large central charge limit for the regimes characterized by $x\approx 0$ and $x\approx 1$ when the intervals are far away and in proximity respectively.\footnote{Note that as explained in \cite{Calabrese:2012nk}, the proximity regime $x\approx 1$ does not involve setting the separation $d$ between the intervals equal to the UV cutoff $\epsilon$ in the \cft{2} with a clear hierarchy $l_1,l_2>>d>\epsilon$. In particular it is not equivalent to the limit $x\to 1$ which will force $d=0$.} These two regimes are described in the $s$ and $t$ channel approximations for the four point correlator involving the fusion of distinct pairs of the twist fields.\footnote{Note that the $s$ and $t$ channels are characterized by $0<x<1/2$ and $1/2<x<1$ respectively.} As shown in \cite{Calabrese:2012nk} the non universal function vanishes non perturbatively in the regime $x\approx 0$ for the $s$ channel at large $c$. On the other hand, for the regime $x\approx 1$ for the $t$ channel, the four point twist correlator admits the following conformal block expansion as described in \cite{Kulaxizi:2014nma}
\begin{equation}\label{block_expand}
\left\langle\mathcal{T}_{n_e}(z_1)\overline{\mathcal{T}}_{n_e}(z_2)\overline{\mathcal{T}}_{n_e}(z_3)
\mathcal{T}_{n_e}(z_4)\right\rangle_\mathbb{C}
=\sum_p\mathcal{F}(c,h_p,h_i,x)\overline{\mathcal{F}}(c,\bar{h}_p,\bar{h}_i,\bar{x}),
\end{equation}
where the summation is taken over all the primary operators with conformal dimensions $(h_p,\bar{h}_p)$, and $(h_i,\bar{h}_i)$ represent the conformal dimensions of the twist operators on the left hand side of \cref{block_expand}. Note that apart from some particular values of the parameters, it is not possible to analytically derive $\mathcal{F}(c,h_p,h_i,x)$. In the semi classical approximation characterized by $c\to\infty$ with $h_p/c,h_i/c$ fixed, the conformal block exponentiates as follows \cite{Belavin:1984vu,1987TMP....73.1088Z}
\begin{equation}\label{block_expo}
\mathcal{F}(c,h_p,h_i,x)\approx\exp\left[-(c/6)f(h_p/c,h_i/c,x)\right].
\end{equation}
The function $f$ in \cref{block_expo} may be computed through the monodromy properties of the solutions of a second order dif{}ferential equation.\footnote{Refer to \cite{Kulaxizi:2014nma} for details of this monodromy technique and related computations.} In the semi classical regime the dominant contribution was shown to arise from the conformal block for the intermediate operator with the lowest conformal dimension in the exchange channel.

For two disjoint intervals in proximity ($x\approx 1$ in the $t$ channel), the relevant intermediate operator with the lowest conformal dimension is $\overline{\mathcal{T}}_{n_e}^2$ \cite{Kulaxizi:2014nma}. Hence the conformal block with the conformal dimension $h_p=h_{\overline{\mathcal{T}}_{n_e}^2}(\equiv\hat{h})$ provides the dominant contribution to the correlator in \cref{block_expand}. Utilizing \cref{block_expo}, the correlator in \cref{block_expand} may then be expressed at large $c$ as\footnote{Note that $h_i=0$ and $\hat{h}=-c/8$ in the replica limit $n_e\to 1$ \cite{Calabrese:2012ew,Calabrese:2012nk,Kulaxizi:2014nma}.}
\begin{equation}\label{corr_f}
\left\langle\mathcal{T}_{n_e}(z_1)\overline{\mathcal{T}}_{n_e}(z_2)\overline{\mathcal{T}}_{n_e}(z_3)
\mathcal{T}_{n_e}(z_4)\right\rangle_\mathbb{C}
\approx\exp\left[-(c/3)f(\hat{h}/c,h_i/c,x)\right],
\end{equation}
and the function $f$ in this case is determined from a monodromy analysis to be \cite{Kulaxizi:2014nma}
\begin{equation}\label{f_x}
f=(3/4)\ln(1-x),
\end{equation}
which describes a universal divergent behavior as $x$ approaches 1. The four point twist correlator in \cref{corr_f} may then be expressed at large $c$ as
\begin{equation}\label{corr_x}
\lim_{n_e\to 1}
\left\langle\mathcal{T}_{n_e}(z_1)\overline{\mathcal{T}}_{n_e}(z_2)\overline{\mathcal{T}}_{n_e}(z_3)
\mathcal{T}_{n_e}(z_4)\right\rangle_\mathbb{C}=(1-x)^{2\hat{h}}.
\end{equation}
The entanglement negativity for two disjoint intervals in proximity ($x\approx 1$) may now be computed from \cref{corr_x} as follows \cite{Kulaxizi:2014nma}
\begin{equation}\label{en_x}
\mathcal{E}=\frac{c}{4}\ln(1-x)
=\frac{c}{4}\ln\left(\frac{\lvert z_{13}\rvert\lvert z_{24}\rvert}{\lvert z_{14}\rvert\lvert z_{23}\rvert}\right),
\end{equation}
where we have utilized the relation $x\equiv (z_{12}z_{34})/(z_{13}z_{24})$. The solution for a general value of the cross ratio $x$ is not amenable to analytic methods, however a consistent numerical analysis has been described in \cite{Kulaxizi:2014nma} although several open issues remain.\footnote{Note that in \cite{Kudler-Flam:2018qjo} the corresponding bulk EWCS has been numerically evaluated assuming an ad hoc conformal block like expansion. As the EWCS is a bulk geometrical quantity, it is not clear from their analysis why such an expansion should be valid.}

We note here that in \cite{Hartman:2013mia}, the entanglement entropy for two disjoint intervals at large $c$ limit was shown to display a phase transition from its $s$ channel value to its $t$ channel value at $x=1/2$. Interestingly in \cite{Kulaxizi:2014nma} a similar phase transition was demonstrated through the numerical analysis described above for the corresponding entanglement negativity at large $c$, from its $s$ channel value of zero to its $t$ channel value given by \cref{en_x}, although the corresponding critical value of $x$ for the transition could not be established. However, due to the existence of a correspondence between the classical geometries dual to the \renyi{} entanglement entropy and the \renyi{} entanglement negativity as shown in \cite{Dong:2018esp}, it could be expected that this phase transition also occurs at $x=1/2$.

\section{HEN and replica symmetry breaking saddle}\label{app_proof}

In this appendix we review a plausible derivation of the holographic proposal involving the algebraic sum of the areas of backreacting cosmic branes as described in \cite{KumarBasak:2020ams} which utilized the replica symmetry breaking saddle reported in \cite{Dong:2021clv}. As discussed in \cite{Calabrese:2012ew,Calabrese:2012nk}, a \renyi{} generalization for the entanglement negativity may be defined as follows
\begin{equation}\label{REN}
\mathcal{N}^{(k)}(A:B)=\text{Tr}\left[\left(\rho_{AB}^{T_B}\right)^k\right]=\frac{Z\left[\mathcal{M}_{2n}^{A,B}\right]}{\left(Z\left[\mathcal{M}_1\right]\right)^{2n}}\,,
\end{equation}
where $\mathcal{M}_{2n}^{A,B}$ denotes the replicated manifold for the entanglement negativity where different copies of the subsystem $A$ are glued cyclically and the copies of $B$ are glued anti-cyclically, $\mathcal{M}_1$ denotes the original manifold and $Z$ denotes their respective path integrals. The entanglement negativity is then given by the analytic continuation of even \renyi{} negativities ($k=2n$) defined above to $k=1$ as
\begin{equation}\label{ENF}
\mathcal{E}(A:B)=\lim_{n\to 1/2}\log\mathcal{N}^{(k=2n)}(A:B).
\end{equation}
In holography this implies that the entanglement negativity at leading order is related to the corresponding bulk gravitational on-shell actions in the saddle point approximation as follows
\begin{equation}\label{PI}
\frac{Z[\mathcal{M}_{2n}^{A,B}]}{(Z[\mathcal{M}_1])^{2n}}=\frac{Z[\mathcal{B}_{2n}^{A,B}]}{(Z[\mathcal{B}_1])^{2n}}=e^{2nI_{\text{grav}}[\mathcal{B}_1]-I_{\text{grav}}[\mathcal{B}_{2n}]}\,.
\end{equation}

Note that the odd analytic continuation of \cref{REN} to $k=1$ does not lead to entanglement negativity but simply gives back the trace condition. Quite interestingly, the authors in \cite{Dong:2021clv} demonstrated that the replica symmetric gravitational saddle is the same for even and odd $k$ and leads to a vanishing result for the entanglement negativity
\begin{equation}
\mathcal{E}^{(\text{sym})}(A:B)=\lim _{k\to 1}\log N_{k}^{(\text{sym})}=0.
\end{equation}
The authors showed that remarkably the dominant saddle corresponding to the entanglement negativity breaks the replica symmetry in the bulk spacetime. The holographic construction of the replica non-symmetric saddle is as follows: Consider $2n$ copies of the bulk manifold which are cut along three non-overlapping codimension-one homology hypersurfaces $\Sigma_A$, $\Sigma_B$ and $\Sigma_{\overline{AB}}$ that obey the homology condition $\partial\Sigma_X=X\cup\gamma_X$, where $\gamma_X$ is the codimension-two hypersurface homologous to X. Now different homology hypersurfaces are glued as follows:
\begin{itemize}
\item $\Sigma_A$: odd numbered copies of the bulk manifold are glued cyclically whereas the even ones are glued to themselves,
\item $\Sigma_B$: even numbered copies of bulk manifold are glued anti-cyclically whereas the odd ones are glued to themselves,
\item $\Sigma_{\overline{AB}}$: all the copies are glued pairwise.
\end{itemize}
Observe that the above construction respects the replica symmetry in the boundary, however it is explicitly broken from $\mathbb{Z}_{2n}\to\mathbb{Z}_n$ in the bulk. This led the authors in \cite{Dong:2021clv} to consider the $\mathbb{Z}_n$ quotient of the original manifold
\begin{equation}
\hat{\mathcal{B}}_{2n}^{A,B(\text{nsym})}=\mathcal{B}_{2n}^{A,B(\text{nsym})}/\mathbb{Z}_n.
\end{equation}
This quotienting has the following effect on the corresponding gravitational on-shell actions
\begin{equation}
I_{\text{grav}}\left[\mathcal{B}_{2n}^{A,B(\text{nsym})}\right]
\equiv nI_{\text{grav}}\left[\hat{\mathcal{B}}_2^{A,B(\text{nsym})}\right]
=nI_{\text{grav}}\left(\mathcal{M}_2^{AB},\gamma_{A_1}^{(n)},\gamma_{B_2}^{(n)}\right),
\end{equation}
where $I_{\text{grav}}(\mathcal{M}_2^{AB},\gamma_{A_1}^{(n)},\gamma_{B_2}^{(n)})$ corresponds to the on-shell bulk action for the alternative construction of the quotiented bulk manifold $\hat{\mathcal{B}}_2^{A,B(\text{nsym})}$ with conical deficits along the codimension-two surfaces $\gamma_{A_1}^{(n)}$ and $\gamma_{B_2}^{(n)}$ (subscripts 1 and 2 simply indicate in which copy the codimension-two surface is located), and having $\mathcal{M}_2^{AB}$ as its asymptotic boundary. Substituting the above equation in \cref{PI} and utilizing the result obtained in \cref{REN} lead to the following expression for the \renyi{} entanglement negativity
\begin{equation}\label{DongF}
\log N_{2n}^{(\text{even},\text{nsym})}
=-n\left[I\left(\mathcal{M}_2^{AB},\gamma_{A_1}^{(n)},\gamma_{B_2}^{(n)}\right)-2I\left(\mathcal{B}_1\right)\right].
\end{equation}
Following the above result, in \cite{KumarBasak:2020ams} the authors utilized a result for the bulk action away from $n=1$ given in \cite{Nakaguchi:2016zqi,Kawabata:2021vyo} to arrive at the following expression for the on-shell action
\begin{equation}
I_{\text{grav}}\left(\mathcal{M}_2^{AB},\gamma_{A_1}^{(n)},\gamma_{B_2}^{(n)}\right)
=2 I_{\text{grav}}\left[\mathcal{B}_1\right]+\frac{\mathcal{A}^{(1/2)}\left(\gamma_{AB}\right)}{4G}
+\left(1-\frac{1}{n}\right)\frac{\mathcal{A}^{(n)}\left(\gamma_A\right)+\mathcal{A}^{(n)}\left(\gamma_B\right)}{4G}.
\end{equation}
Implementing the limit $n\to 1/2$ in the above equation, inserting it into \cref{DongF}, and finally substituting the result obtained into \cref{ENF} lead to the following expression for the holographic entanglement negativity
\begin{align}
\mathcal{E}(A:B)&=\left.\frac{1}{8G_N}\middle[\mathcal{A}^{(1/2)}\left(\gamma_A\right)
+\mathcal{A}^{(1/2)}\left(\gamma_B\right)-\mathcal{A}^{(1/2)}\left(\gamma_{AB}\right)\right]\\
&=\frac{1}{2}I^{(1/2)}(A:B),\label{HMIEN}
\end{align}
where in the last line $I^{(1/2)}(A:B)$ denotes the holographic \renyi{} mutual information of order half for the bipartite system $AB$. Note that in order to arrive at \cref{DongF}, the authors in \cite{Dong:2021clv} assumed that the subsystems $A$, $B$ and $AB$ are together in a tripartite pure state. Upon imposing this assumption in the holographic constructions for the entanglement negativity of a single interval, two adjacent intervals, and two disjoint intervals \cite{Chaturvedi:2016rcn,Jain:2017aqk,Malvimat:2018txq}, the authors in \cite{KumarBasak:2020ams} demonstrated that they all reduce to the above expression.

\section{HEN and Markov gap}\label{app_gap}

In this appendix we describe the crucial role of the holographic Markov gap to the holographic entanglement negativity proposals described in \cite{Chaturvedi:2016rft,Malvimat:2018txq,Kudler-Flam:2018qjo,Kusuki:2019zsp}. Before we discuss the relation of the Markov gap to the holographic entanglement negativity, let us briefly review the definition of the reflected entropy described in \cite{Dutta:2019gen} and the holographic Markov gap as explained in \cite{Hayden:2021gno}. To this end consider a bipartite mixed state $\rho_{AB}$. There exists a canonical purification $\lvert\sqrt{\rho_{AB}}\rangle$ in the doubled Hilbert spaces $\mathcal{H}_A\otimes\mathcal{H}_B\otimes\mathcal{H}_{A^*}\otimes\mathcal{H}_{B^*}$. The von Neumann entropy of the bipartite state $\rho_{AA^*}$ which is obtained from the state $\lvert\sqrt{\rho_{AB}}\rangle$ by tracing over the degrees of freedom of $BB^*$ is known as the reflected entropy \cite{Dutta:2019gen}
\begin{align}
S_R(A:B)&=S_{AA^*}=-\text{Tr}(\rho_{AA^*}\log\rho_{AA^*}),\\
\rho_{AA^*}&=\text{Tr}_{BB^*}( \lvert\sqrt{\rho_{AB}}\rangle\langle\sqrt{\rho_{AB}}\rvert).
\end{align}
Now consider the action of a quantum channel $\mathcal{R}_{B\to BC}$ which acts on the bipartite mixed state $\rho_{AB}$ to produce a tripartite state $\tilde{\rho}_{ABC}$
\begin{equation}
\tilde{\rho}_{ABC}=\mathcal{R}_{B\to BC}\left(\rho_{AB}\right).
\end{equation}
The Markov recovery process refers to reproducing the quantum state $\rho_{ABC}$ whose reduction led to the bipartite mixed state $\rho_{AB}$ through the operation of the above quantum channel $\mathcal{R}$ which acts on the subsystem $B$ alone. If $\tilde{\rho}_{ABC}=\rho_{ABC}$, the Markov recovery process is said to be perfect and the state $\rho_{ABC}$ is said to be a quantum Markov chain for the ordering $A\to B\to C$. Furthermore, it has been shown through quantum information techniques that this happens when the conditional mutual information $I(A:C\mid B)$ vanishes \cite{Petz:1986tvy}. This result was further refined in \cite{2015CMaPh.340..575F} where the authors demonstrated that there exists a bound expressed as follows
\begin{equation}
I(A:C\mid B)\geq-\max_{\mathcal{R}_{B\to BC}}\log F\left(\rho_{ABC},\mathcal{R}_{B\to BC}\left(\rho_{AB}\right)\right),
\end{equation}
where $F(\rho_{ABC},\mathcal{R}_{B\to BC}(\rho_{AB}))$ is the quantum fidelity which is unity when $\tilde{\rho}_{ABC}=\rho_{ABC}$, and zero when the two density matrices have support on orthogonal subspaces. This led the authors to examine the above bound for the Markov recovery process of the reduced density matrix $\rho_{ABB^*}$ that occurs in the canonical purification $\rho_{ABB^*}=\text{Tr}_{A^*}(\lvert\sqrt{\rho_{AB}}\rangle\langle\sqrt{\rho_{AB}}\rvert)$. This immediately implies the following constraint on the conditional mutual information because of the above inequality
\begin{align}
I(A:B\mid B^*)&\geq-\max_{\mathcal{R}_{B\to BB^*}}\log F\left(\rho_{ABB^*},\mathcal{R}_{B\to BB^*}\left(\rho_{A B}\right)\right),\\
S_R(A:B)-I(A:B)&\geq-\max_{\mathcal{R}_{B\to BB^*}}\log F\left(\rho_{ABB^*},\mathcal{R}_{B\to BB^*}\left(\rho_{A B}\right)\right),
\end{align}
where in the last line the conditional mutual information $I(A:B\mid B^*)$ is simply re-expressed in terms of the reflected entropy and the mutual information which the authors in \cite{Hayden:2021gno} termed as the Markov gap. Furthermore, in the context of \ads{3}/\cft{2} the authors demonstrated that the above bound may be expressed geometrically as follows
\begin{equation}\label{MG}
S_{R}(A:B)-I(A:B)\geq\frac{\ell_{\text{AdS}}(\log 2)}{2G_N}
\times(\#\text{ of boundaries of EWCS})+\mathcal{O}\left(\frac{1}{G_N}\right),
\end{equation}
where $\ell_{\text{AdS}}$ refers to the \ads{} radius, $\#$ of boundaries of EWCS denotes the number of end points of EWCS in the bulk \ads{3} geometry (boundaries at asymptotic infinity are not considered as they are infinitely far away).

In order to understand the connection of the holographic Markov gap to the holographic entanglement negativity consider an alternative proposal for it developed in \cite{Kudler-Flam:2018qjo,Kusuki:2019zsp}. As explained in detail in \cref{sn_HEN_EWCS}, the authors in \cite{Kudler-Flam:2018qjo} proposed that the holographic entanglement negativity is given by the area of the backreacting EWCS. This proposal was further refined in \cite{Kusuki:2019zsp} where the authors related it to the \renyi{} reflected entropy of order half as follows
\begin{align}
\mathcal{E}_0&=\frac{1}{2}S_{R}^{(1/2)}(A:B)\\&=\frac{\mathcal{X}_d}{2}S_{R}(A:B),
\end{align}
where the second equality is valid when $AA^*$ and $BB^*$ share a spherical entangling surface in higher dimensions or when $AA^*$ is a continuous interval in \ads{3}/\cft{2} \cite{Kusuki:2019zsp}. We have denoted the entanglement negativity computed from this proposal as $\mathcal{E}_0$ to distinguish it from that obtained through the proposal involving the mutual information of order half in \cref{HMIEN}. Consider now the difference between the holographic entanglement negativity computed using these two proposals
\begin{equation}
\mathcal{E}-\mathcal{E}_0=\left.\frac{1}{2}\middle[S_{R}^{(1/2)}(A:B)-I^{(1/2)}(A:B)\right].
\end{equation}
As discussed in \cref{HREE}, for subsystems with spherical entangling surfaces and in \ads{3}/\cft{2}, the \renyi{} entropies of order half of $A$, $B$ and $AB$ are proportional to their corresponding von Neumann entropies as given by \cref{SAchi} and hence the above equation reduces to
\begin{align}
\mathcal{E}-\mathcal{E}_0&=\left.\frac{\mathcal{X}_d}{2}\middle[S_{R}(A: B)-I(A:B)\right],\\
\mathcal{E}-\mathcal{E}_0&\geq\frac{3\ell_{\text{AdS}}(\log 2)}{8G_N}
\times(\#\text{ of boundaries of EWCS})+\mathcal{O}\left(\frac{1}{G_N}\right),
\end{align}
where in order to arrive at the last line of the above equation we have utilized the inequality given in \cref{MG} and $\mathcal{X}_2=3/2$. Thus the difference between the entanglement negativities computed from the two proposals is proportional to the Markov gap which is non-vanishing in a holographic \cft{2} as described by the above equation.

%%%%%
\bibliographystyle{JHEP}
\bibliography{references}
%%%%%
\end{document}